\documentstyle[12pt,moriond]{article}
\newcommand{\HI}{\mbox{H\,{\sc i}}}
\newcommand{\mhi}{\mbox{${\cal M}_{HI}$}}
\newcommand{\msun}{\mbox{${\cal M}_\odot$}}
\newcommand{\lsun}{\mbox{${\cal L}_\odot$}}
\newcommand{\kms}{\mbox{km s$^{-1}$}}
\newcommand{\dark}{$\frac{{\cal M}_{HI}}{L_{B}}$}
\newcommand{\am}[2]{$#1'\,\hspace{-1.7mm}.\hspace{.0mm}#2$}
\newcommand{\as}[2]{$#1''\,\hspace{-1.7mm}.\hspace{.0mm}#2$}

\begin{document}
\heading{PROPERTIES OF `SUPERTHIN' GALAXIES}

\centerline{L.D. Matthews $^{1}$, W. van Driel $^{2}$, J.S. Gallagher$^{3}$}
\centerline{$^{1}$National Radio Astronomy Observatory, 
Charlottesville, U.S.A.} 
\centerline{$^{2}$Unit\'e Scientifique Nan\c{c}ay, Observatoire de Paris, France} 
\centerline{$^{3}$Astronomy Department, University of Madison-Wisconsin, U.S.A.}

\begin{moriondabstract}
`Superthins' are a subset of edge-on spiral galaxies exhibiting disks 
with large axial ratios (a/b$>$10), extraordinarily
small stellar scale heights, and no bulge component. These are thus
examples of dynamically cold, pure disk galaxies.
We have recently obtained multiwavelength (\HI, optical, NIR) observations 
for a large sample of superthin spirals. Our data lead to a picture of
superthins as `underevolved' disk galaxies in both a dynamical and a 
star formation sense. Studies of these relatively simple disk systems can
therefore provide unique constraints on galaxy disk formation and
evolution without looking beyond the local universe.
We present the results of a detailed analysis of the 
nearby superthin UGC 7321 as an illustration of these ideas. 
\end{moriondabstract}

\section{Meet the galaxies}
Vorontsov-Vel'yaminov (1967) was one of the first to draw attention 
to a  unique subset of edge-on spiral galaxies that exhibit extraordinarily 
large disk axial ratios and no discernible bulge component. Goad \&
Roberts  (1981) dubbed these galaxies `superthins' and recognized 
that spirals selected on the basis of their superthin
morphologies tend to share other unique properties, including low 
optical surface brightness disks, high neutral gas fractions, 
low metallicities, and slowly rising, dwarf-like rotation curves (see also
Karachentsev \& Xu 1991; Bergvall \& R\"onnback 1995; Matthews {\it et al.} 
1999a). 

Superthins are by no means uncommon in nearby space. 
From an inspection of photographic
survey plates, Karachentsev {\it et al.} (1993) compiled
a catalogue of 4454 edge-on, pure disk galaxies (the Flat Galaxy
Catalogue or FGC). Our group has surveyed 474 of the FGC objects in the
\HI\ 21-cm line using the Nan\c{c}ay Radio Telescope and Green Bank
140-ft Telescope (Matthews \& van Driel, in preparation). 
We detected over 50\% of our targets within $V_{h}<9500$~km~s$^{-1}$
(see also Giovanelli {\it et al.} 1997).
The high detection rate of our survey demonstrates
that optically organized, gas-rich, pure disk galaxies are abundant
in the nearby universe, and hence represent one of the most common
products of galaxy disk formation.
Galaxy formation paradigms must therefore explain the
abundance of these small disks and their unique properties, even if their
contribution to the overall matter density of the universe is small.

We have  obtained follow-up optical imaging and photometry of 95 of our
\HI-detected galaxies using the WIYN telescope at Kitt Peak. 
By definition, all of the galaxies we surveyed have large 
disk axial ratios and little or no bulge component. However,
we find the FGC galaxies are nonetheless morphologically diverse; many are 
true superthin objects with  small stellar scale heights, while a 
number exhibit thicker, more flocculent disks. Interestingly, galaxies
of both types may have similar disk rotational velocities, \HI\ contents, 
and optical luminosities.

\begin{figure}
\includegraphics{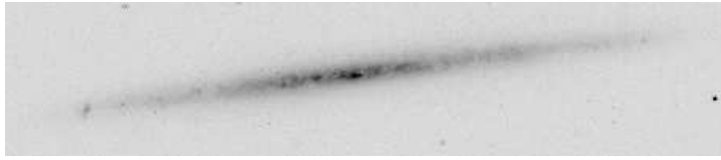}  
\vspace{25mm}
\caption{$R$-band CCD image of UGC~7321. Field size is $\sim$\am{5}{5} across.}
\end{figure}

The objects exhibiting superthin morphologies are of particular interest, since
the thinness of their stellar disks implies they are among the least dynamically
evolved of nearby disk galaxies. Our optical images reveal that the superthins 
frequently appear rather diffuse, indicating the stellar densities in their 
disks are low. Since corresponding
\HI\ contents are generally high, this suggests that these  galaxies have 
been inefficient star-formers. Thus superthins can be viewed as highly
`underevolved' systems. For this reason, superthins 
can offer us a glimpse of the conditions during the 
early stages of quiescent disk galaxy  evolution without
looking beyond the local universe. Moreover, these simple disks allow
us to probe disk structure and dynamics without the complication of a
bulge or large internal extinction. As an illustration of these ideas, 
we summarize the results from a detailed analysis of the nearby superthin 
UGC~7321.

\section{A nearby superthin studied in detail: UGC 7321}
\subsection{Global properties}
Located at a distance of $\sim$10Mpc, UGC~7321  is a  prototypical 
example of a superthin spiral (Fig.~1). 
Using the WIYN telescope, we obtained photometrically-calibrated $B$
and $R$ images of UGC~7321 under \as{0}{6} seeing conditions. 
In addition, we obtained complementary
NIR $H$-band imaging with IRIM on the Kitt Peak 2.1-m telescope,
an \HI\ pencil-beam map using the Nan\c{c}ay Telescope, and we have analyzed
archival VLA \HI\ observations of this galaxy. For more details on the 
observations and their analysis, we refer the reader to Matthews {\it et al.} 
(1999a).

For UGC~7321 we derive the following global 
properties\footnote{All quantities have been corrected for Galactic
and internal extinction and projected to a face-on value; 
see Matthews {\it et al.} (1999a).}: $i\approx88^{\circ}$;
$M_{B}=-17.0$; $\overline{\mu}_{B}=27.6$~mag~arcsec$^{-2}$; 
$A_{opt}$=16.3~kpc; \mhi=1.1$\times10^{9}$\msun; \dark=1.1 \msun/\lsun;
$W_{20}$=233 \kms; $h_{r}$=2.1~kpc.
We find the rotation curve of UGC~7321 rises slowly, and begins to
flatten only well outside the stellar disk. UGC~7321 can therefore
be characterized as a low surface brightness,
gas-rich galaxy with a rather weak central mass concentration. 
Its global  properties are thus in some ways more reminiscent of an 
Irregular galaxy than an Sd spiral. Nonetheless,
the  stellar disk of UGC~7321 is clearly highly organized, and
its double-horned global \HI\ profile  is distinctly spiralesque.  

\subsection{The radial light distribution}
A fit to its azimuthally-averaged brightness distribution 
reveals that UGC~7321 does not have a simple exponential stellar
disk (Fig.~2). At small radii
a brightness excess over the best exponential fit is observed, while at
large radii, the light profile falls off faster than predicted for an
exponential, suggesting the stellar disk of UGC~7321 may be truncated.
In addition, a major axis brightness profile extracted from our $H$-band data 
exhibits distinct `steps' in the light distribution.
These observations suggest that perhaps viscous evolution has been inefficient
in this low-density galaxy (see also Matthews \& Gallagher 1997). This
also provides evidence against the hypothesis that the exponential nature 
of the {\it stellar} disk of spirals is established by the initial conditions 
of galaxy formation (cf. Dalcanton {\it et al.} 1997).

\subsection{Disk colors and color  gradients}
The examination of disk color gradients offers an important means of
constraining disk formation mechanisms (e.g., de Jong 1996) and 
dynamical evolution histories (e.g., Just {\it et al.} 1996). UGC~7321
is particularly suited to the exploration of disk color gradients,
since it suffers minimally from internal dust extinction
(Matthews {\it et al.} 1999a).

Near the center of its disk, UGC~7321 exhibits a small, very red
 nuclear feature ($B-R\approx$1.5), only a few arcseconds
across. Surrounding this feature is a more extended red region with
$B-R\sim$1.2, which extends to $r\approx\pm20''$ on either side of the
disk center, and has a rather distinct boundary. Intriguingly, this
region corresponds very closely to the region over which we observe a
light excess over a pure exponential disk (see above). Similar regions
have also been found in 3 other superthins (Matthews {\it et al.} 1999b); 
perhaps these represent ancient starbursts,
the cores of the original protogalaxies, or kinematically distinct
subsystems analogous to the bulges of other spirals.
Cutting into the red central region of UGC~7321, we find thin blue
bands of stars along the midplane of the galaxy. These bands
grow both thicker and bluer with increasing galactocentric radius,
having $B-R\approx$1.05 at $|r|=20''$, and 
reaching $B-R\sim$0.5 at the visible edges of the disk. This radial
bluing cannot be explained solely by dust, and is consistent with
the type of color gradient predicted by `inside-out' galaxy formation
models (e.g., White \& Frenk 1991). 
A faint, thicker, but highly flattened disk of unresolved stars is also
visible surrounding the UGC~7321 disk at $|r|\le$2.0$'$. This
component has $B-R\approx$1.1, and shows little change in color with
galactocentric distance. The color of this component is consistent
with a population of `old disk' stars. Their location at higher
$z$-heights than the bluer stars along the disk midplane implies that
some dynamical heating has occurred even in dynamically cold superthins.

\begin{figure}
\includegraphics{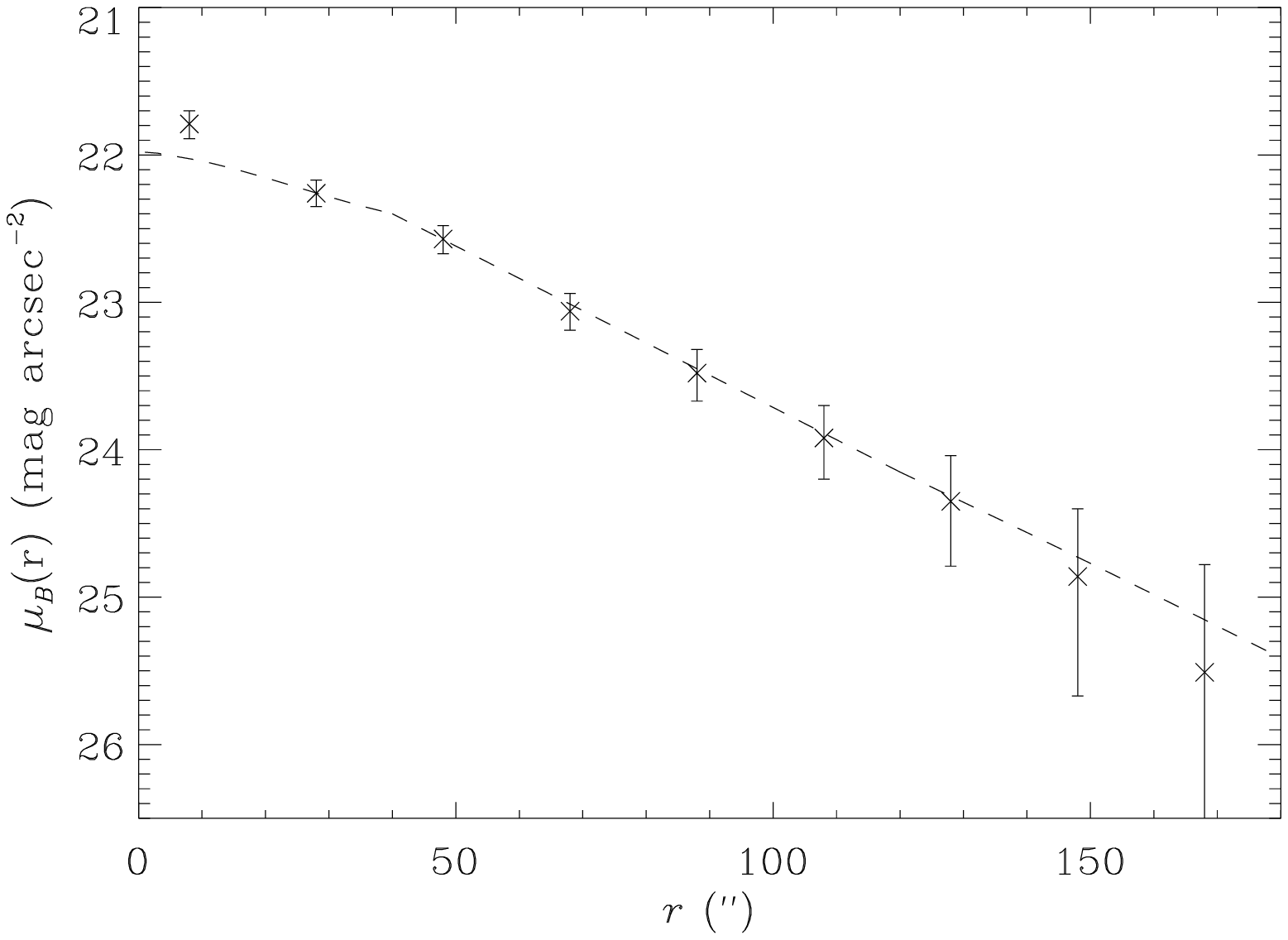}  
\includegraphics{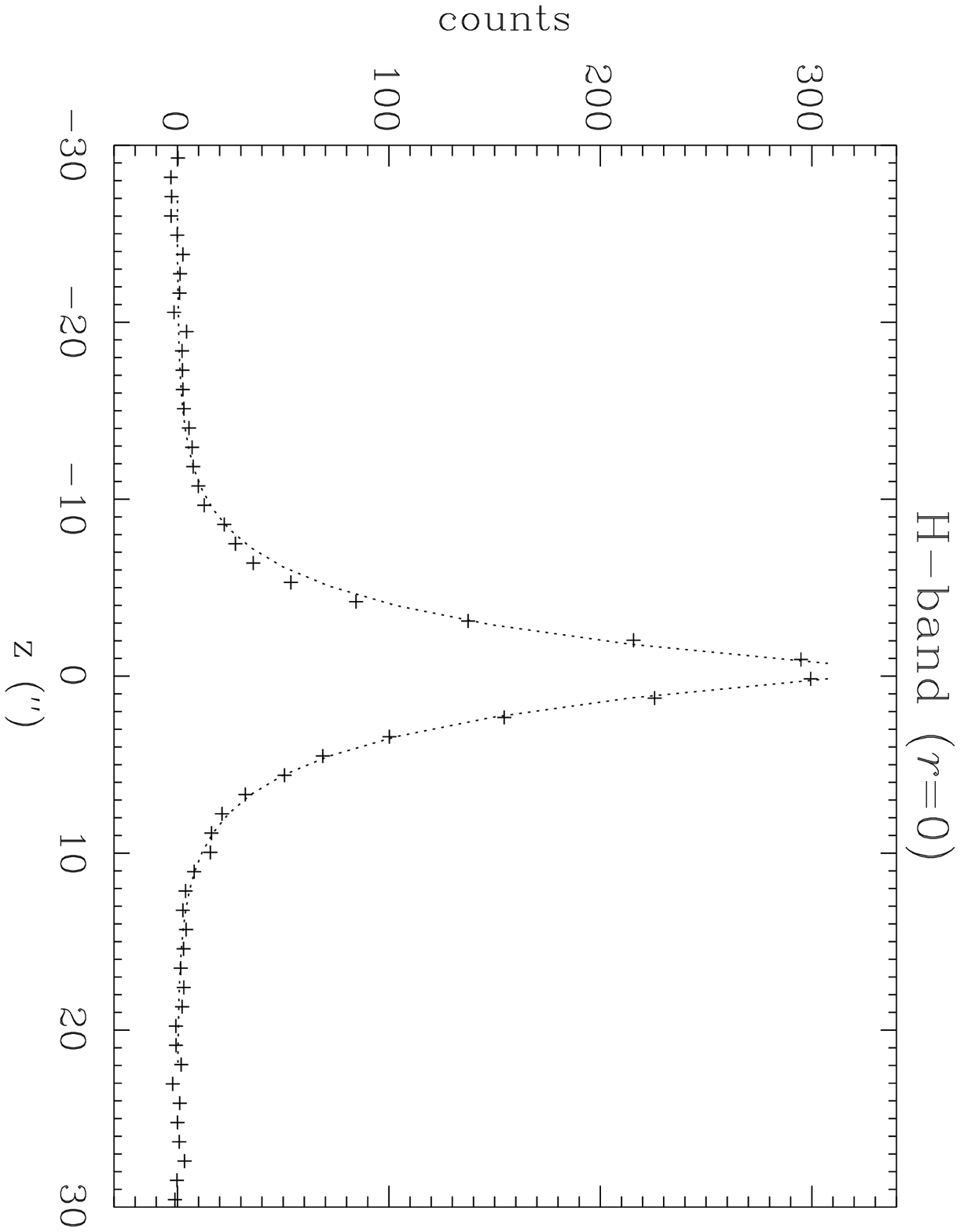}
\vspace{40mm}
\caption{(Left) Azimuthally-averaged radial $B$-band surface brightness 
profile of UGC~7321 with best-fitting exponential overplotted as a dashed
line. Figure 3: (Right) Vertical $H$-band brightness
profile of UGC~7321 along the disk minor axis, with the best-fitting
exponential overplotted as a dotted line.}
\end{figure}

We note that the outer disk regions of UGC~7321 are too blue
to be explained by low metallicity alone and must be quite
young. Nonetheless, the simultaneous presence of stars with $B-R>1$
implies that UGC~7321 is not a young galaxy, but rather one which has
evolved very slowly. This is contrary to the picture of blue,
gas-rich, low surface brightness galaxies as young systems (see also 
Jimenez {\it et al.} 1998).  

\subsection{Vertical disk structure}
Measurements of the vertical light profiles of galaxy disks provide
insight into their formation, stability, and evolutionary histories
(e.g., de Grijs 1997 and references therein).
In order to characterize the vertical light distribution of UGC~7321,
we have performed functional fits to the brightness profiles extracted 
at various galactocentric radii from our $H$- and $R$-band images.
We find the disk of UGC~7321 is not locally isothermal over most of its
radial extent. At $r$=0, the vertical light profile can be well characterized 
by a single exponential function with a scale height $h_{z,c}\approx$140~pc 
(Fig~3). At intermediate galactic radii (\am{0}{5}$\ge|r|\ge$\am{1}{5}), 
the vertical light profile becomes less
peaked than an exponential and can be represented as the
sum of two `sech' functions of differing scale heights 
($h_{z,2}\approx$120~pc and $h_{z,3}\approx$218~pc). For
$|r|\ge$\am{1}{5} we cannot rule out that the disk may be
approximately isothermal.

We interpret these results as evidence for the existence of 
disk subcomponents in UGC~7321, analogous to the disk subcomponents
of the Milky Way (MW; e.g., Freeman 1993). At intermediate galactic
radii, the two fitted sech functions may represent
 components similar to the ``young
disk'' and the ``thin disk'' of the MW, while near the disk center,
the exponential nature of the brightness profile suggests the
existence of an additional component of extremely small scale height,
perhaps analogous to the MW's ``nuclear disk''.
This multi-disk interpretation appears
consistent with 
the existence of various disk subpopulations delineated in our
color maps (see above). Even apparently simple disks like the
superthins thus appear to be quite structurally complex and to have been
subject to some degree of dynamical evolution.

\
\vfill
\end{document}